\documentclass[titlepage]{article}
\usepackage[utf8]{inputenc}
\usepackage[pdf]{graphviz}
\usepackage{float}
\usepackage{xcolor}
\usepackage[a4paper, total={6in, 8in}]{geometry}
\usepackage[sorting=none, style=numeric, backend=biber, date=year, isbn=false, url=false,eprint=false]{biblatex} 
\usepackage{amsmath}
\usepackage{bm}
\usepackage{soul}
\usepackage{caption}
\usepackage{lineno}
\usepackage[affil-it]{authblk}

\newcommand{\pkg}[1]{{\normalfont\fontseries{b}\selectfont #1}} \let\proglang=\textsf 
\graphicspath{{./fig/}}
\newcommand{\hlcite}[1]{\colorbox{yellow}{#1}}
\renewcommand\hl[1]{#1} 
\renewcommand\hlcite[1]{#1}

\title{Evaluating the probative value of forensic gait analysis evidence using empirical data}
\author[1,*]{Ruoyun Hui}
\author[1,2]{Amy L Wilson}
\author[2]{Colin Aitken}
\author[3]{Ivan Birch}
\author[3]{Nadia Asgeirsdottir}
\author[4,5]{Graham Jackson}
\affil[1]{\small The Alan Turing Institute, London NW1 2DB, UK}
\affil[2]{School of Mathematics and Maxwell Institute of Mathematical Sciences, University of Edinburgh, Edinburgh EH9 3FD, UK}
\affil[3]{FGA Services, UK}
\affil[4]{Advance Forensic Science, St. Andrews KY16 0NA, UK}
\affil[5]{School of Applied Sciences, Division of Psychology and Forensic Science, Abertay University, Dundee DD1 1HG, UK}
\affil[*]{Corresponding author: Ruoyun Hui, rhui@turing.ac.uk}

\begin{document}
\maketitle
\begin{abstract}
    Forensic gait analysis can aid the investigation of crimes through comparing features of gait captured in video footage. Modelling the probative value of gait evidence requires an understanding of the variation of features of gait between individuals in the population and within the same individuals. We address this question using a previously described population dataset and newly collected datasets with repeated observations of the same individuals on separate occasions. In addition to exploring the level of variability, correlation between features of gait, and the effect of demographic factors, we developed a likelihood ratio model through recoding features of gait as dichotomous variables and dimension reduction using PCA. \hl{High correlations between some features were observed, confirming that they should not contribute independently to the weight of evidence.} The likelihood ratio model produced misleading likelihood ratios in less than 10\% of the comparisons using the first four principal components. However, the risk increases when within-individual variability is mis-specified. \hl{Therefore, while the current model provides assistance to the judgement of gait experts, }human expertise is indispensable to decide whether or not the difference in walking and/or recording conditions between the reference and questioned footage could have caused any observed differences in the features of gait. We discuss future directions in understanding the sources of the variability, improving statistical modelling \hl{and note the need to consider carefully how to select the relevant population for model fitting.}
    \\
    
\end{abstract}

\newpage
\section{Introduction}

There is a long history of using features of gait to assist the investigation of crime \cite{nirenberg_review_2018}. In recent years, gait evidence captured on CCTV footage has been increasingly used as forensic evidence. The irregularities around the quality of video footage, lighting, camera angles and other conditions that might differ between the footage of known and unknown source make it challenging to automate the process. Forensic gait analysis as currently practised therefore primarily relies on manual annotation of features of gait and subjective interpretation of evidence.  

As many forensic fields are undergoing a paradigm shift towards a more probabilistic framework \hlcite{\cite{FSR_LR,ICCA, primer,european_network_of_forensic_science_institutes_enfsi_2016}}, forensic gait analysis has come under criticism \cite{abboud_forensic_2017,van_mastrigt_critical_2018,macoveciuc_forensic_2019}. The main points raised have included 1) gait is not unique to each individual, thus cannot be used for identification; 2) the performance of gait experts and their methods has not been validated; 3) there is a lack of understanding on the variation of features of gait backed by empirical data. However, it has never been claimed in the context of forensic gait analysis based on video footage, that gait is unique, only that it can contribute to the process of identification. Competency testing has been available for a number of years for practitioners in the UK, US and Canada, and there is now a published and widely used method\hl{, the Sheffield Features of Gait Tool} \cite{birch_repeatability_2019,birch_accuracy_2021}\hl{, which is based on human-derived observations of gait}. Much progress has been made in recent years to establish the empirical basis for observational gait analysis, including \hl{research on} the development and validation of \hl{this} features of gait tool \cite{birch_development_2013}, measurement of inter- and intra-observer errors \cite{birch_repeatability_2019,GroningenMethod,seckiner_forensic_2024}, establishment of population databases \cite{birch_aiding_2016, seckiner_forensic_2024}, and exploration of factors affecting the observed features of gait \cite{wiedemeijer_effects_2018,reidy_effect_2020,seckiner_development_2021,bergsma_foot_2021,GroningenMethod}. Intra- and inter-subject variability of features of gait, as used in forensic gait analysis, is now the \hl{topic} of a number of studies, and is a key element of this paper. \hl{The three criticisms given at the beginning of the paragraph}, therefore, do not constitute a valid reason to \hl{completely} dismiss the potential value of gait evidence. 

Central to a scientific approach is the need to account for and communicate any uncertainties using transparent models backed by empirical data. Following the Bayesian reasoning framework, the likelihood ratio (LR) has been \hl{widely accepted in academic fields and is moving towards acceptance} across many forensic fields as a logical and balanced way to report the probative value of evidence \cite{evett_logical_2015,european_network_of_forensic_science_institutes_enfsi_2016,MORRISON2022100270}. The evidence in forensic gait analysis typically consists of the questioned footage of an unknown individual at the crime scene and the reference footage of a known individual at a different location. In this context, an LR is defined as the ratio between the probability of \hl{obtaining the observed} features of gait in the reference and questioned footage if the prosecution proposition (usually, that the same person appeared in the reference and the questioned footage) is true, and the probability of \hl{obtaining the observed} features of gait in the reference and questioned footage if the defence proposition (usually, that different people were captured in the reference and the questioned footage) is true:

\[
LR = \frac{Pr(\bm{y_1}, \bm{y_2} | H_p)}{Pr(\bm{y_1}, \bm{y_2} | H_d)}
\]

where $\bm{y_1}$ and $\bm{y_2}$ represent \hl{observations} in the reference and questioned footage; $H_p$ and $H_d$ represent the prosecution proposition and the defence proposition. This formulation makes it clear that the expert is concerned solely with assigning probabilities for the gait observations, conditioned on a pair of competing propositions; the expert does not consider the probability of the truth of the competing propositions.

The literature on how to arrive at these probabilities from observable features of gait is scarce. The \hl{variations} that need to be \hl{taken into account} to evaluate an LR can arise from many sources: the same person might walk differently on separate occasions, either due to intrinsic variations or affected by extrinsic factors such as the terrain or incline; the filming environment and the quality of the footage can make some features difficult to observe. \hl{Note also that variability in gait is not the same as measured variability in gait which may include observer error.} Jackson and Birch \hlcite{\cite{jackson_probative_2020}} described a conceptual framework for gait analysts to assign subjective probabilities for the numerator and the denominator of the LR. Otten and Wiedemeijer developed an LR model using categorical features of gait by performing PCA on the polychoric correlation matrix, followed by kernel density functions fitted for each PC \cite{GroningenMethod}. \hl{Variabilities in the assessments by the experts} were measured and incorporated into their model, although the underlying features of each individual are assumed to be invariable. Using a different method to extract features of gait, Seckiner \cite{seckiner_development_2021} \hl{presented a model that calculated LRs for individual features that are deemed independent from each other, then multiplied them to produce the final LR}. As well as the loss of information arising from not including the correlated features, the model does not account for within-individual variations between recording sessions. It is not clear from the existing literature whether or not within individual variations in gait affect the outcomes of observational gait analysis in the forensic context, or if they do, to what extent. It is also not clear how any such effects should be accounted for within any statistical framework for assessing gait evidence.

In this study we first explore the between-individual variation of the features of gait in a population database \cite{birch_aiding_2016} and the within-individual variation observed in two newly-collected datasets. Then, we build on these foundations to develop a statistical model inspired by \cite{GroningenMethod} to evaluate the probative value of gait analysis evidence that can account for within-individual variation in gait. We also use our datasets to explore the range of likelihood ratios and to measure the calibration of the system. The sensitivity to individual data points and assumptions about within-individual variation highlights gaps in the empirical data for further research to improve the model. \hl{Due to these data limitations, and the other qualitative uncertainties discussed above that are inherent in current approaches for forensic gait analysis, we do not suggest that our model is ready for immediate use in casework. Instead, it should be seen as a first step for the assessment of likelihood ratios for forensic gait analysis.}

\section{Datasets}
In observational gait analysis, the features of gait are noted by analysts after watching a number of gait cycles, taking into consideration \hl{limitations} related to the quality of the footage, camera angles, and other environmental factors. The features are annotated as categorical variables. The data used in this study are collected specially for forensic gait analysis; therefore features have been selected according to their relevance in forensic case work, rather than clinical practice \cite{birch_aiding_2016}.

\hl{Several aspects of features of gait might affect their use in forensic gait analysis.} These include:

\begin{itemize}
    \item The prevalence of features of gait in the population (i.e. variations between individuals)
    \item The dependency between features of gait
    \item The variation in features of gait within individuals
    \item The differences in variability between individuals 
    \item The effect on features of gait of intrinsic and extrinsic \hl{(e.g. surface, footwear)} factors
\end{itemize}

In this article, we explore the first two aspects using a population database, and the next two aspects using two small datasets of repeated walks by the same subjects. Our current study did not collect data on the last aspect, but we briefly discuss how the effects might be incorporated in statistical modelling.  

\subsection{Variations between individuals}

A population database has been created to assist forensic gait analysis \cite{birch_aiding_2016}. It consists of \hl{observations of 1,007 members of the public} made across seven geographical locations in the UK by one experienced gait analyst. Sixteen features of gait (some recorded for both the left and right side) along with the estimated sex, age group, weight, height, and ethnicity were recorded for each individual. Pedestrians exhibiting factors that may substantially \hl{affect} their gait (e.g. carrying bags, or holding hands with another person) were excluded.

The demographic profile of the database has been presented in \cite{birch_aiding_2016}. Here we explore the effects of demographic factors on features of gait, an understanding of which is instrumental when assessing the relevancy of a reference population \cite{champod_establishing_2004}. We also examine the correlation between different features of gait in the population. 

\subsubsection{Association of features of gait with demographic factors}

\begin{figure}
    \centering
    \includegraphics[width=\textwidth]{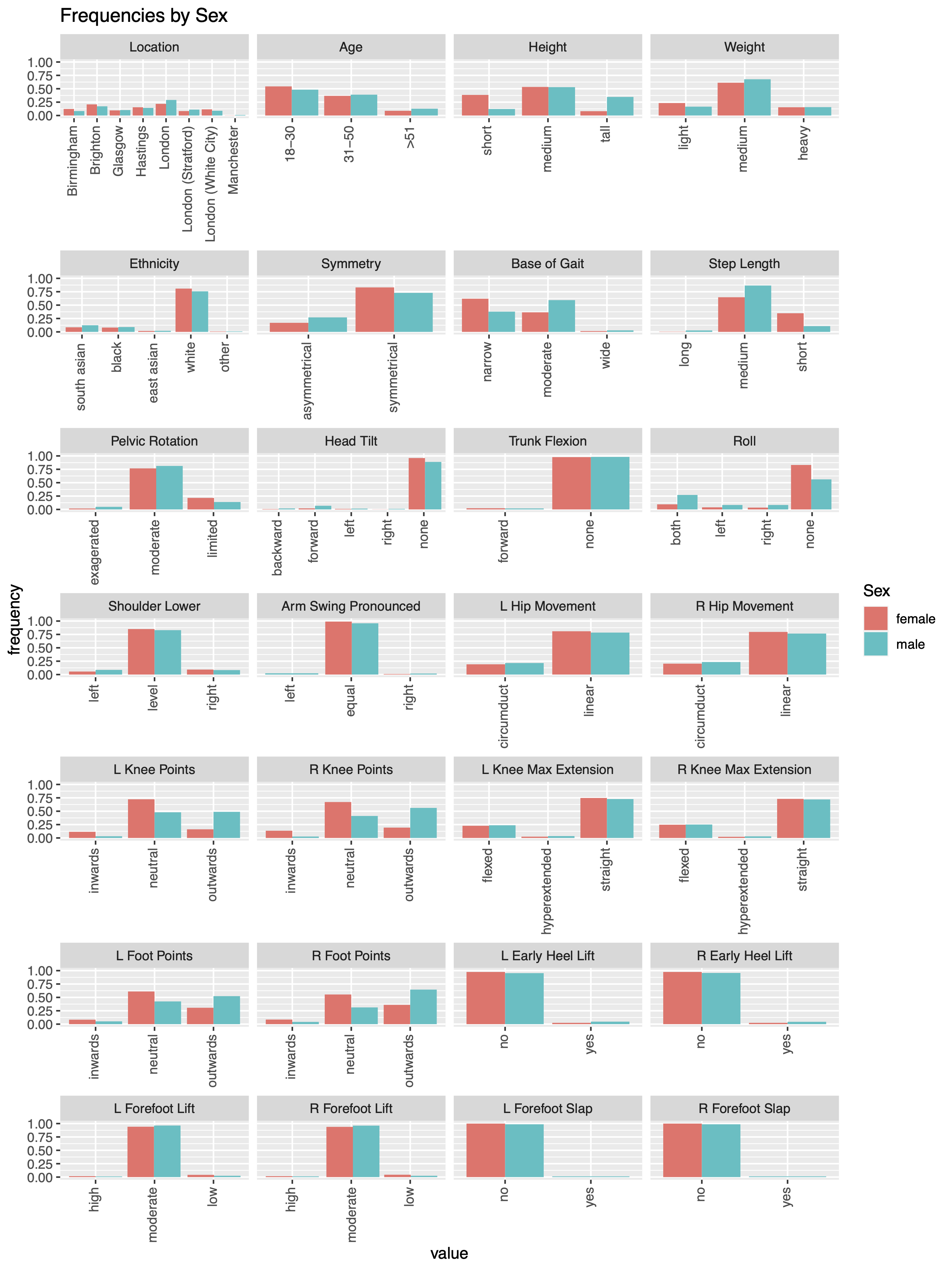}
    \caption{Distribution of features of gait in the population database by sex}
    \label{fig:database_by_sex}
\end{figure}

Features of gait may vary with body proportions, sex, and age; additionally, the effect of cultural factors may show up through ethnicity \cite{connor_biometric_2018} or location. For example, Figure \ref{fig:database_by_sex} shows the distribution of other demographic factors and features of gait grouped by sex. In this dataset, female pedestrians \hl{seem to }have narrower base of gait, take shorter steps, show less rolling of the head, and keep their knees and feet neutral or pointing inward more often compared to the male pedestrians \hl{, but these differences require more research}. We note that ethnicity and ancestry have much more nuance than the categories in the database; regrettably the data collection process required the expert to make a quick judgement from appearance only. 

We fitted a series of logistic regression models to explore the association, using the features of gait as response variables, and demographic factors as explanatory variables. For features with two possible states ($Y_i \in \{0, 1\}$) such as the symmetry of gait, the binary logistic model is used:

\[
\ln \frac{Pr(Y_i = 1)}{Pr(Y_i = 0)} = \ln \frac{Pr(Y_i = 1)}{1-Pr(Y_i = 1)} = \alpha + \bm{\beta \cdot X_i} 
\]

where $\bm{X_i}$ is the vector of the demographic variables for individual $i$, $\alpha$ is a constant and $\bm{\beta}$ is the vector of coefficients.

For features with more than two possible states, we first checked that they followed a meaningful order, then performed ordinal logistic regression. Two features were each split into two new variables to ensure a meaningful order: head tilt (none, forward, backward, left, right) into frontal head tilt (left, none, right) and sagittal head tilt (backward, none, forward); roll (left, right, none, both) into roll left (yes, no) and roll right (yes, no).  

In ordinal logistic regression, assuming there are $L$ ordered states of the response variable ($Y_i \in \{0, 1, ...  L-1\}, L \geq 3$), the log odds of exceeding each threshold $l$ is modelled as:

\[
\ln \frac{Pr(Y_i > l)}{Pr(Y_i \leq l)} = \ln \frac{Pr(Y_i > l)}{1-Pr(Y_i > l)} = \alpha_l + \bm{\beta \cdot X_i}, l = 1, ...  L-1
\]

The ordered states $\{0, 1, ... L-1\}$ do not carry numerical significance. 

For each feature, we first fit two models to assess the influence of ethnicity and the location of recording, respectively. After controlling for the biological variables (sex, height, weight, and age), if the coefficient of any of the four indicator variables about ethnicity is statistically significant, they were included in the final model; similarly with the seven indicator variables about the location. The biological variables were always included in the final model since they are widely acknowledged to have a possible influence on gait. 

Supplementary Table 1 summarises the regression coefficients in the final model. They show the average effect of each variable on the odds ratio on the log scale. The intercepts $\alpha$ and $\alpha_l$ are related to the base rates of the states and not informative about the influence of the explanatory variables. Note that the significance probability ($p$-value) by itself is not an indicator of the magnitude of influence. The results suggest that sex, height, weight, and age all affect features of gait, but the relative importance differs between different features of gait. After controlling for them, the influence of ethnicity is limited. The location of observation, however, sometimes stands out regarding the frequency of certain features of gait: for example the asymmetrical gait was observed more often in Birmingham, less often in White City (in London) and in Hastings; in Brighton more people lift their heels early. Perhaps the population being observed differ in other unmeasured aspects between the cities, or between the sites of observation - the population present at an airport is likely to have a different makeup compared to the population at a shopping centre, for example. We note accordingly that what constitutes the relevant population for a given case would have bearings on the weight of the gait evidence. 

\subsubsection{Correlation between features of gait}

\begin{figure}
    \centering
    \includegraphics[width=\textwidth]{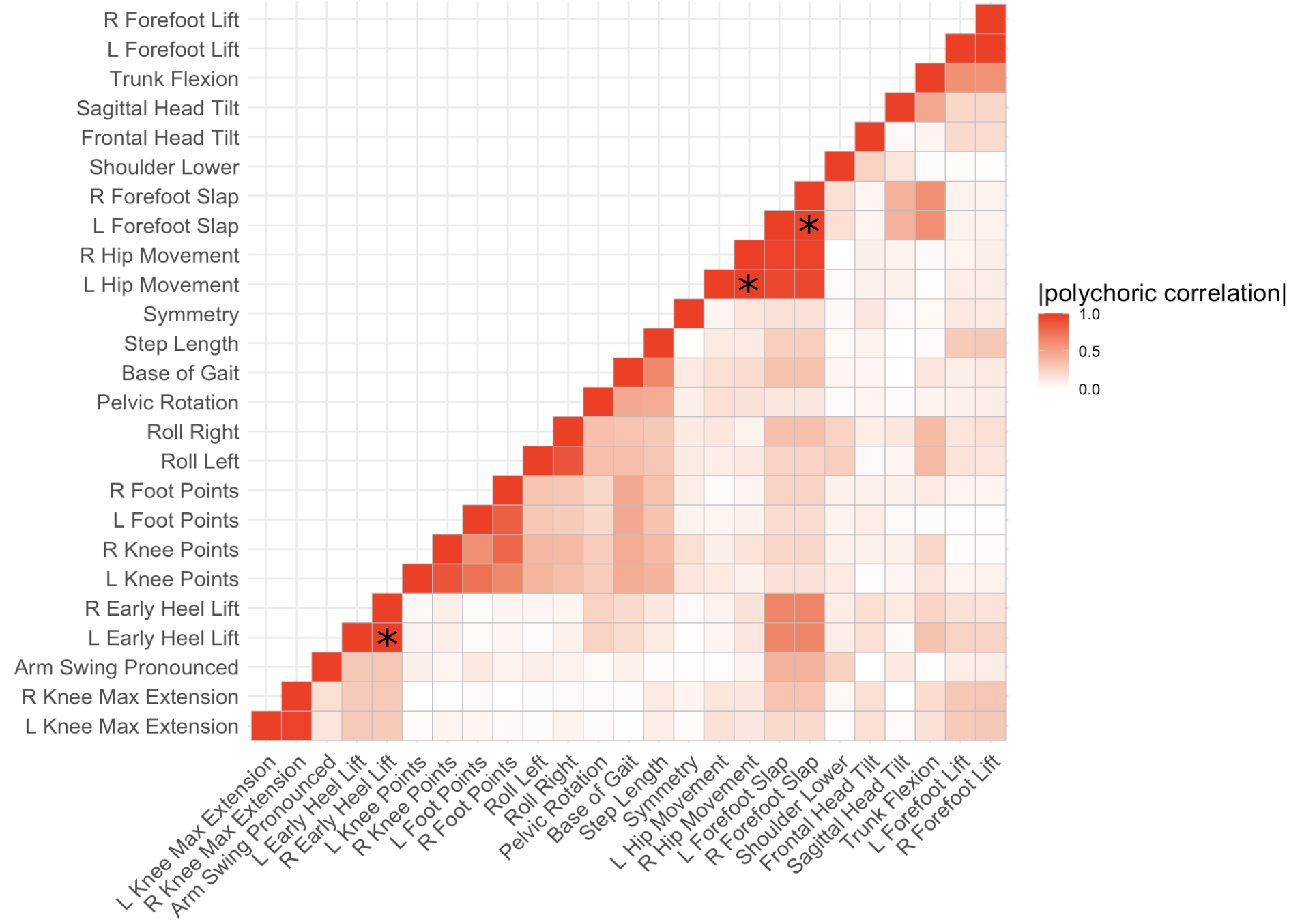}
    \caption{Polychoric correlation between features of gait in the population database by sex. The absolute values are shown considering the arbitrary direction of ordering. *: the high agreement of hip movement, early heel lift and forefoot slap on the left/right side caused numerical problems during computation; 0.9999 was assigned to these pairs.}
    \label{fig:correlation}
\end{figure}

Features of gait are expected to show correlations considering the biomechanical constraints of the human body. Figure \ref{fig:correlation} shows the polychoric correlation coefficients between all features of gait recorded in the population database, after splitting head tilt and roll into new variables to ensure a meaningful ordering. This procedure assumes that each ordinal variable is controlled by an underlying continuous variable that follows a normal distribution, and produces the maximum likelihood estimate of the correlation coefficient between these continuous variables \cite{pearson1900}. Because the direction of the ordering is arbitrary, the absolute values of the polychoric correlation coefficients were plotted. For features recorded separately for the left and the right side, the correlation between them is usually very high (in particular, extremely high correspondence between left and right early heel lift, hip movement, and forefoot slap causes numerical problems when estimating the polychoric correlation). The direction in which the knee points is also correlated with the direction of the foot (0.7-0.8 on the same side). None of the people showing forefoot slap in this dataset has a forward trunk flexion, hence those two features appear correlated. Similarly, forefoot slap and early heel lift never occur in the same person.

The observation confirms that features of gait are not independent from each other. The strength of correlation estimated from empirical data might assist gait experts in the subjective interpretation of gait evidence. It also shows that a LR model needs to account for the correlation between the features of gait. 

\subsection{Variations within individuals}\label{within}
To date there are no empirical data on how the features of gait used in observational gait analysis change in the same person on different occasions. Ideally this should be jointly modelled with between-individual variations in a population database; the setup of population data collection, however, prevents repeated observations of the same individuals. As part of this study, we collected two sets of data. The first was collected from \textit{ad hoc} self-recorded footage of the researchers', under uncontrolled conditions using no protocol. The second was collected from controlled footage of 18 participants as part of structured study into variations in gait. 

\subsubsection{Dataset A}
\begin{figure}
    \centering
    \includegraphics[width=\textwidth]{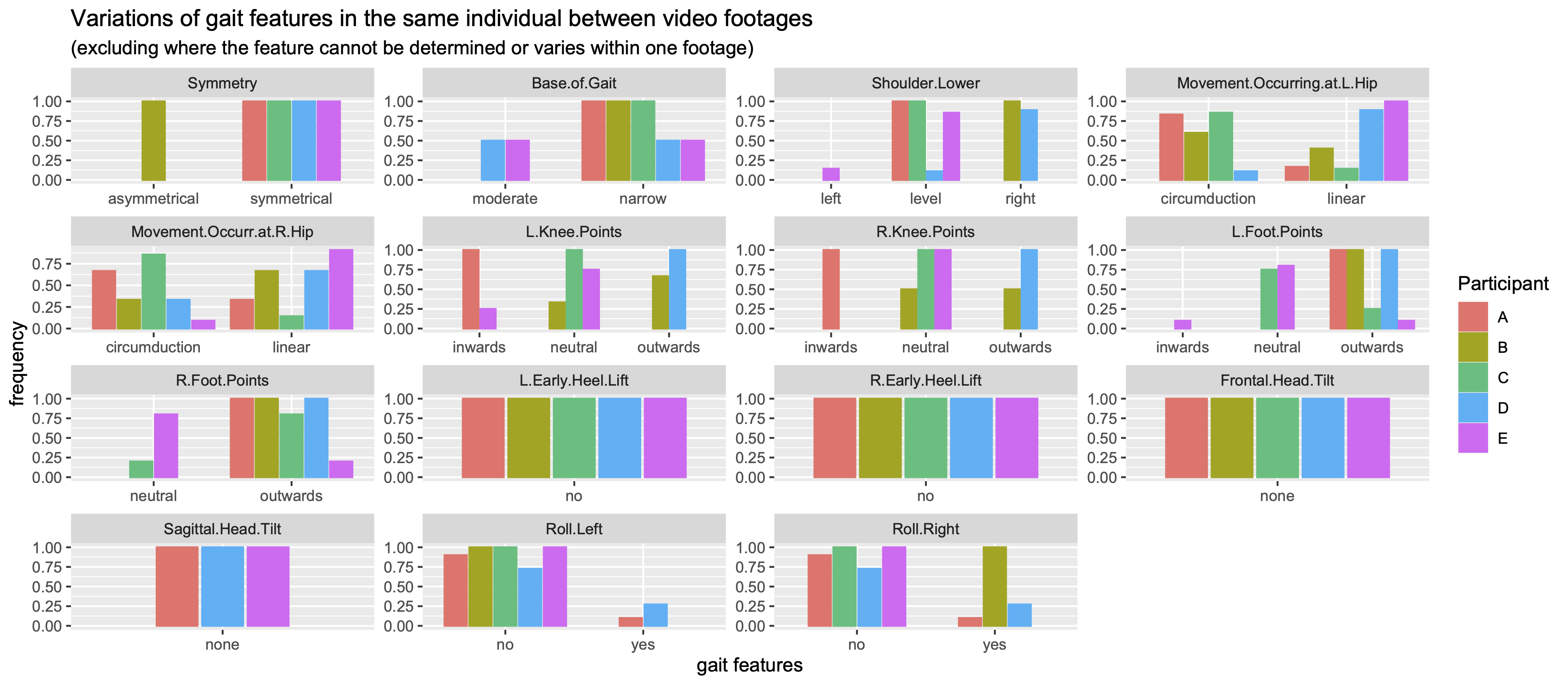}
    \caption{Distribution of features of gait in footage from Dataset A, after removing instances where each feature cannot be determined or show variations within one walk. Only features that also appear in the population database are shown.}
    \label{fig:within-ind-studyA}
\end{figure}

Five of the authors and their colleagues used mobile devices to record themselves walking on a number of separate occasions, varying from 6 to 11. No protocol was implemented and location, terrain, footwear, clothing and perspective varied both between occasions and participants, thus maximizing the potential for variation in both gait \hl{and} the observation of features of gait. One gait expert \hlcite{(same as in \cite{birch_accuracy_2021})} analysed all the footage. After removing instances where a particular feature could not be determined, the distribution of the states of the features of gait in each participant is shown in Figure \ref{fig:within-ind-studyA} (only showing features of gait that also appear in the population database, although the forensic gait analysis tool has since been expanded). For consistency with the population database, only one state was chosen for each feature of gait on each occasion, although the gait tool allows more nuance to be recorded (for example, an individual might vary in a feature of gait between steps). We note that taking the most common state exhibited on each occasion could exaggerate the within-individual variation, if that variability between steps is an intrinsic part of the individual's gait.

The objective of this data collection was to produce a dataset that was subject to a wide range of possible variation as the result of a range of intrinsic and extrinsic factors. As anticipated, the dataset does demonstrate variation in the observed features of gait, and the degree of variation appears to vary both between subjects and between features of gait.

The dataset highlights the complexity of the interaction between the various factors that can affect gait. However, the lack of consistent conditions also caused a good deal of missing data (for example, when features could not be determined), and made it impossible to disentangle the intrinsic biological variations from the effect of extrinsic factors (such as terrain and line of progression) and recording settings. A second dataset \hl{(Dataset B)} was therefore collected from recruited participants in more controlled conditions using an experimental protocol.

\subsubsection{Dataset B}

The aim of this study was to understand variations in features of gait under controlled conditions. Participants were recruited to attend three recording sessions on consecutive Mondays. During the recording session, after giving consent to participate in the study and to share their anonymised features of gait, each participant was asked to walk naturally on flat ground for 20 metres, turn back, and walk back to the starting point. The route was designed to include an extra segment leading up to the starting point to help participants relax into their natural gait. Participants were asked to attend in the same footwear for all recording sessions, and to remove bulky outerwear and any objects in hand during the walk, although there were still instances of negligence. Two cameras on tripods recorded the gait from frontal and sagittal plane perspectives. The study location was changed after the first recording session because the ground was found to be slightly sloped.

Eighteen participants completed the recording on all three days. The footage was analysed by the same gait expert after being anonymised and arranged in random order. In contrast to the results from the researchers' self-recording, we found remarkably limited variation within the participants: there were only 8 instances where the same individual displayed more than one state of a feature of gait across the three days, among 15 features of gait (that overlap with the population database) annotated for 18 participants (Supplementary Table 2). 

\section{Evaluation of the weight of evidence}
\subsection{Likelihood ratio model}
As a first attempt at producing a quantitative likelihood ratio from categorical features of gait, we consider a two-level model for multivariate data that incorporates both between-source variation and within-source variation \cite{aitken_evaluation_2004}:

\begin{equation} \label{eq:1}
    LR = \frac{\int_\theta f(\bm{y_1} | \theta, U) f(\bm{y_2}|\theta, U) f(\theta) d \theta} {\int_\theta\{f(\bm{y_1}|\theta, U)f(\theta)\}d\theta \int_\theta \{f(\bm{y_2}|\theta, U) f(\theta)\}d\theta}
\end{equation}

Here $\bm{y_1}$ and $\bm{y_2}$ refer to the reference and questioned evidence; the function $f(\theta)$ describes between-individual variation, namely the distribution of individual mean $\theta$ across the population; $U$ describes the variance within an individual, so that each observation $y$ is drawn according to $f(y|\theta, U)$.

However, one major obstacle is that features of gait are ordinal variables and also not independent of each other. Building on work by \cite{GroningenMethod} ("Groningen method"), we performed Principal Components Analysis (PCA) to transform the ordinal data matrix into new independent variables. The main difference is instead of using the polychoric correlation matrix of ordinal variables in PCA, we transformed the ordinal variables to binary variables first, followed by regular PCA. Since the polychoric correlation comes from the underlying continuous variables, we consider it inappropriate to treat the ordinal indices as numeric to obtain scores on the principal components. Although PCA is most suitable for continuous data, it is equivalent to multidimensional scaling using the pairwise Euclidean distance when applied on binary variables \cite{jolliffe_principal_1986}. An alternative approach is to perform PCA from the polychoric correlation matrix, but use the expected value of the underlying continuous variable conditioned on the ordinal level to obtain the projected scores \cite{kolenikov_use_2004}. This approach, however, produced higher rates of misleading evidence in our analysis using the dataset with high within-individual variability (Supplementary Figure 1). 

\begin{figure}
    \centering
    \includegraphics[width=\textwidth]{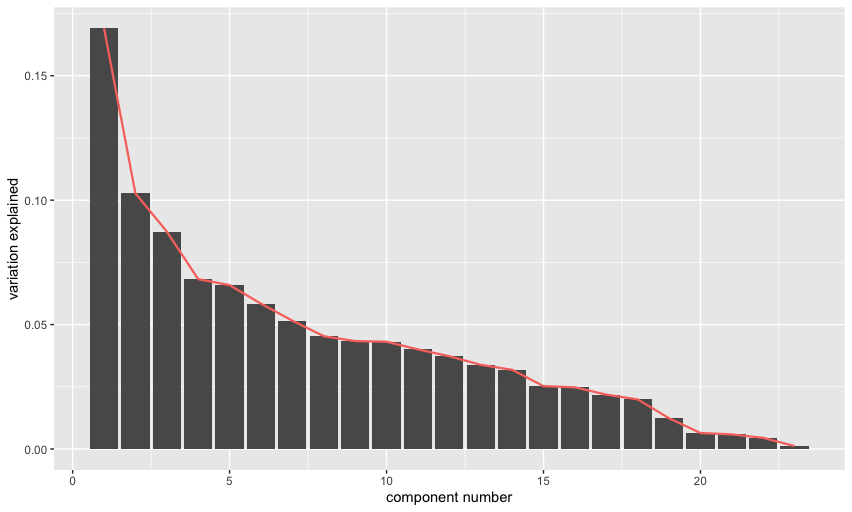}
    \caption{The amount of variation explained by each principal component in the population dataset recoded into binary variables.}
    \label{fig:scree_plot}
\end{figure}

To transform the ordinal variables, a feature of gait $j$ with $L_j$ ordered levels \{0, 1, ..., $L_j - 1$\} can be recoded into $L_j - 1$ binary variables:

\begin{align*}
& b_{ik} = 
\begin{cases}
    1 & \text{if } c_j > k \\
    0 & \text{otherwise}
\end{cases}
\end{align*}

where $c_j$ is the original value of the variable, $k \in \{0, 1, ..., L_j - 2\}$. The background population data $\bm{Z}$, reference footage data $\bm{Y_1}$ (from known persons of interest) and questioned footage data $\bm{Y_2}$ (from the crime scene) were all transformed in this way into binary variable forms $\bm{Z_b}$, $\bm{Y_{1b}}$, and $\bm{Y_{2b}}$. Then we performed PCA (using the \pkg{psych}\cite{r_psych} package in \proglang{R}\cite{R}) on the new binary population data $\bm{Z_b}$. The result of PCA establishes a transformed space by constructing new variables (principal component, or PC) using linear combinations of the binary variables. Figure \ref{fig:scree_plot} shows the amount of total variation in the data explained by each principal component. 

If only the top $M$ principal components are kept, the same truncated linear transformation $\bm{T_M}$ from the PCA can then be applied to $\bm{Z_b}$, $\bm{Y_{1b}}$, and $\bm{Y_{2b}}$ to construct $M$ new variables from $r$ binary variables:
\begin{align*}
\bm{Z^*} &= \bm{Z_b}\bm{T_M} \\
\bm{Y_1^*} &= \bm{Y_{1b}}\bm{T_M} \\
\bm{Y_2^*} &= \bm{Y_{2b}}\bm{T_M} 
\end{align*}

\begin{figure}
    \centering
    \includegraphics[width=\textwidth]{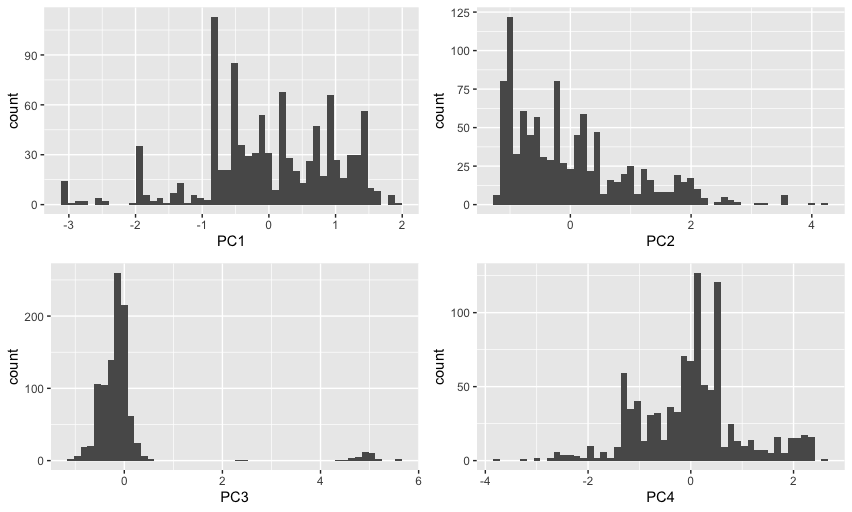}
    \caption{Histogram showing the distribution of the scores on the first four PC using the population data.}
    \label{fig:hist_pc}
\end{figure}

We can then apply the two-level model to $\bm{Z^*}$, $\bm{Y_1^*}$ and $\bm{Y_2^*}$ rather than to the full datasets. The advantage of using PCA to transform the data in this way is twofold. Firstly, we can reduce the number of dimensions that need to be modelled, and secondly, the variables in the transformed dataset are independent to one another so we do not need to model correlations in the data. Figure \ref{fig:hist_pc} shows the distribution of the background population data ($\bm{Z^*}$) for the first four PCs. Because the shape of these distributions is highly irregular, we used a kernel density function for each PC $j (1 \leq j \leq 4)$ applied to the background population data for between-individual variation, i.e. $f(\theta)$ in Equation \ref{eq:1}:
\[
f_j(\theta) = \frac{1}{nh_j}\sum_{i=1}^{n}K\left(\frac{\theta_ - \bar{z_{ij}^*}}{h_j}\right)
\]
where $\theta$ is the parameter representing individual means; $K$ is a Gaussian kernel; $h$ is the smoothing parameter (bandwidth) chosen according to \cite{silverman_choosing_1978} and $\bar{z_{ij}^*}$ is the mean score of individual $i$ on the $j$th PC in the background population dataset. In practice we had to replace $\bar{z_{ij}^*}$ with $z_{ij}^*$, since the population dataset we currently have only includes one observation for each individual. A kernel density function is a non-parametric method of smoothing the data, by taking a weighted average of available data points according to how close they are in terms of the input features to the new point of interest. In this way we do not have to choose a specific probability distribution for the between-individual variation. 

Since our data regarding within-individual variation are scarce, we assume the same amount of variation in each individual (i.e. the same $U$ in the term for within-individual variation, $f(y | \theta, U)$, in Equation \ref{eq:1}). \hlcite{See Section \ref{misspecify} for a discussion of this assumption.} Further, we assume the observations from the reference footage $\bm{Y_1^*} = \{y_{1j}^*\}$ and those from the questioned footage $\bm{Y_2^*} = \{y_{2j}^*\}$ both follow a normal distribution around the individual mean $\theta_{ij}$:
\[
y_{ij}^* \sim \mathcal{N}(\theta_{ij}, s_j^2), i \in \{1, 2\}
\]
The within-individual variance on the $j$th PC $s_j^2$ can be estimated from a background dataset $\bm{Z}$ containing $C_i$ observations of individual $i$. If the dataset consists of $m=\sum_{i=1}^n C_i$ observations across $n$ individuals, 
\[
s_j^2 = \frac{n}{m(m-n)}\sum_{i = 1}^n C_i\sum_{k=1}^{C_i} (z_{kj}^* - \bar{z}_{ij}^*)^2
\]
where $\bar{z}_{ij}^*$ is the average of $C_i$ scores from individual $i$ on the $j$th PC. We discuss later in this paper the issues that arise from lack of data on within-individual variation. With more data it would be possible to check our assumption that each individual has the same variation and to revise this assumption if necessary. 

As the PCs are orthogonal to each other, each PC can be considered independently in a univariate model. The likelihood ratio can then be formed from the product of the likelihood ratios for each PC. Then the likelihood ratio from the first $M$ PC is given by:
\begin{align*}
LR_{M} &= \frac{Pr(\bm{Y_1^*}, \bm{Y_2^*} | H_p)}{Pr(\bm{Y_1^*}, \bm{Y_2^*} | H_d)} \\
&= \frac{Pr(\bm{Y_1^*}, \bm{Y_2^*} | H_p)}{Pr(\bm{Y_1^*} | H_d) \cdot Pr(\bm{Y_2^*} | H_d)} \\
&=\prod_{j=1}^M \frac{\int_\theta f(y_{1j}^*|\theta, s_j) f(y_{2j}^*|\theta, s_j)f_j(\theta)d\theta}{\int_{\theta} f(y_{1j}^*|\theta, s_j) f_j(\theta) d\theta \int_{\theta} f(y_{2j}^*|\theta, s_j) f_j(\theta) d\theta}
\end{align*} 

\subsection{Results}

\subsubsection{Within-individual variance}
We applied the likelihood ratio model to the available datasets. Since our population dataset contains \hl{only} one observation for each individual, the within-individual variation (in the form of the variance of the scores along each PC) has to be estimated from other datasets. \hl{The population database provided the background population data} \hlcite{($Z$)} \hl{and PCA decomposition} \hlcite{($T_M$)}\hl{; reference and questioned footage data} \hlcite{($Y_1$ and $Y_2$)} \hl{were drawn from Dataset B.} We also referred to Dataset A as the variance arising from uncontrolled conditions can be used as an upper bound of within-individual variability. A gait analyst is likely to forgo comparison if the walking and/or recording conditions vary substantially between the reference and the questioned footage in current practice. However, we would like to explore how mis-specifying within-individual variability might affect the performance of our model. \hlcite{This issue is discussed further in section \ref{misspecify}.}

Missing values in these two datasets were replaced by random samples from non-missing observations of the same individual or, if that feature was not observed for that individual at all, they were replaced by random samples from all other individuals in the study. \hl{This will have a small impact on any correlations between features, but as the number of missing values was small the impact should be negligible.} Table \ref{tab:within_var} shows the within-individual variance of the PC scores estimated from both studies. To reduce the bias from filling missing observations as described above, we repeated the process 5,000 times with different random seeds and report the means of the estimates from the 5,000 replicates. Changing the random seed causes the estimate to vary with a standard deviation around 10\% of the mean in Dataset A, and less than 1\% of the mean in Dataset B (due to much less missing data). We chose to use up to the top four PCs because this is the point where the variation starts to level off (Figure \ref{fig:scree_plot}). As expected, apart from PC3 the results from Dataset B show much reduced within-individual variation. \hl{Dataset A in general shows higher within-individual variation for most features, likely due to uncontrolled condition (clothing, terrain, speed), more individual repetitions (6-11, so that variations are more likely to be captured) and more missing data. On PC3, the loadings are concentrated on left and right early heel lift, which happen to show variation within one walker in Dataset B but within none in Dataset A.}

\begin{table}[h]
\centering
\caption{Within-individual variance of the PC scores from two datasets.}
\begin{tabular}{lllll}
\hline
            & PC1   & PC2   & PC3   & PC4   \\ \hline
Dataset A (high variability) & 0.113 & 0.263 & 0.022 & 0.517 \\
Dataset B (low variability) & 0.007 & 0.010 & 0.119 & 0.033 \\ \hline
\end{tabular}
\label{tab:within_var}
\end{table}

\subsubsection{Range of LRs}\label{range}
We also used the features of gait in Dataset B (after filling in missing values) to see how the LR model behaves with real world data. As effectively the only input from Dataset B to the model is the within-individual variance (PCA was performed using the population dataset), we considered it acceptable to use it for model validation. Later, we also attempt using the within-individual variance estimated from Dataset A (Section \ref{misspecify}).

\begin{figure}[h]
    \centering
    \includegraphics[width=\textwidth]{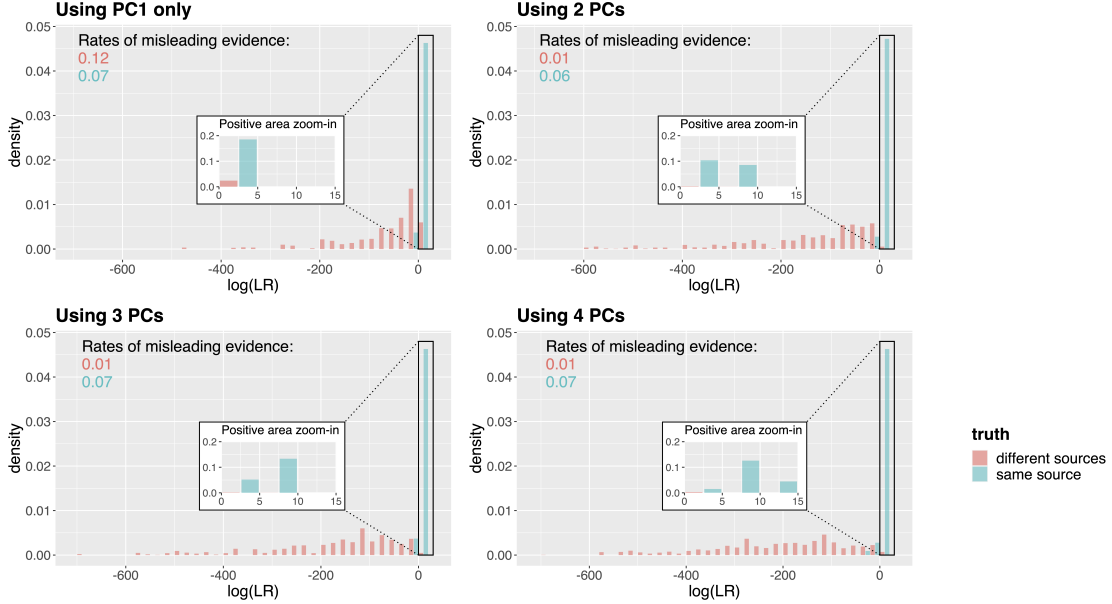}
    \caption{Histogram of log (base e) likelihood ratios obtained using a two-level model from scores on up to four PCs using footage from Dataset B, grouped by the ground truth. The rates of misleading evidence are the proportion of LRs being greater than one in different-source comparisons (pink), and the proportion of LRs being less than one in same-source comparisons (blue). The insets display the positive values on a different scale.}
    \label{fig:lr_hist}
\end{figure}

After one piece of footage was chosen as the questioned footage, other footage from the same individual was treated as the reference footage for same-source comparisons, and all footage from a different individual within the same dataset were treated as reference footage for different-source comparisons. In each dataset, all possible combinations of question and reference footage were used to generate the distribution of LRs grouped by known ground truths (Figure \ref{fig:lr_hist}). The results were further grouped by the total number of PCs used.

The magnitude of the LRs is highly asymmetrical: comparisons between different-source observations can produce log LRs below a few hundreds in the negative, while the largest log LRs from same-source comparisons are around 10. When only the first PC is used, if the questioned footage and reference footage capture different individuals, there is a probability of 0.12 that the model produces a misleading positive log LR (LR $>$1) that supports the same-source proposition over the different-source one; if the questioned footage and reference footage capture the same individual, the probability of the model producing a misleading negative log LR (LR $<$ 1) is 0.07. Adding information from the second PC shifts the probability mass from different-source comparisons in the correct direction and reduces the rate of misleading evidence to approximately 1\%. The gain from adding the third and fourth PC, however, is not as obvious. We therefore focus on the first two PCs in the rest of this section. \hl{Note that some of these LR values are very low, too low to be realistically reported. These LRs are low because the dataset used to calculate the within individual variation is small and was collected under tightly controlled conditions and hence may underestimate the true within individual variation. We discuss this issue further in} \hlcite{Section \ref{misspecify}.} \hl{As we do not have additional data to refine the estimation of this variance it is recommended that in such instances, the reported value be truncated at}\hlcite{$10^{-8}$.}

\subsubsection{Calibration}

\begin{figure}[H]
    \centering
    \includegraphics[width=\textwidth]{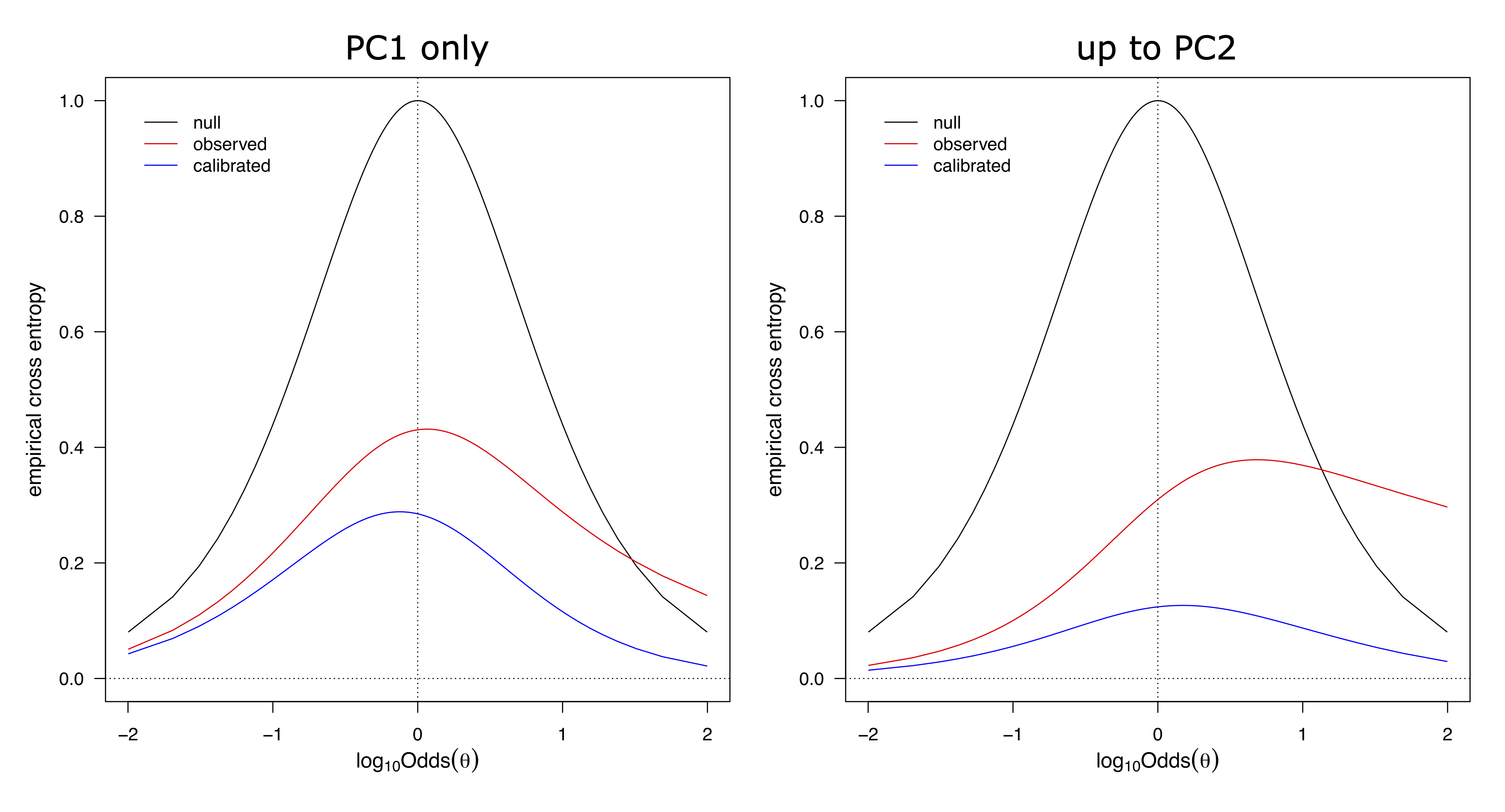}
    \caption{Empirical cross entropy plots of the likelihood ratios produced by the two-level model from scores on one or two PCs using footage from Dataset B.}
    \label{fig:ece}
\end{figure}

To check how well the LRs are calibrated, we produced the empirical cross entropy (ECE) plot \cite{ramos_informationtheoretical_2013} (Figure \ref{fig:ece}). When the prior odds is high (base 10 log odds greater than approximately 1.5), the observed curve rises above the curve from the null model, meaning that our model produces more cross entropy loss than a system that always produces an LR of 1. The loss at large prior odds becomes even worse when the first two PCs are used. \hl{Tippett plots are shown in Supplementary Figure 2. As discussed in}\hlcite{Section \ref{range},}\hl{ some of the LRs for the different source comparisons are very low due to the small dataset used to estimate the within individual variation. We investigate the sensitivity to this variance assumption in the following section.}

\begin{figure}[H]
    \centering
    \includegraphics[width=\textwidth]{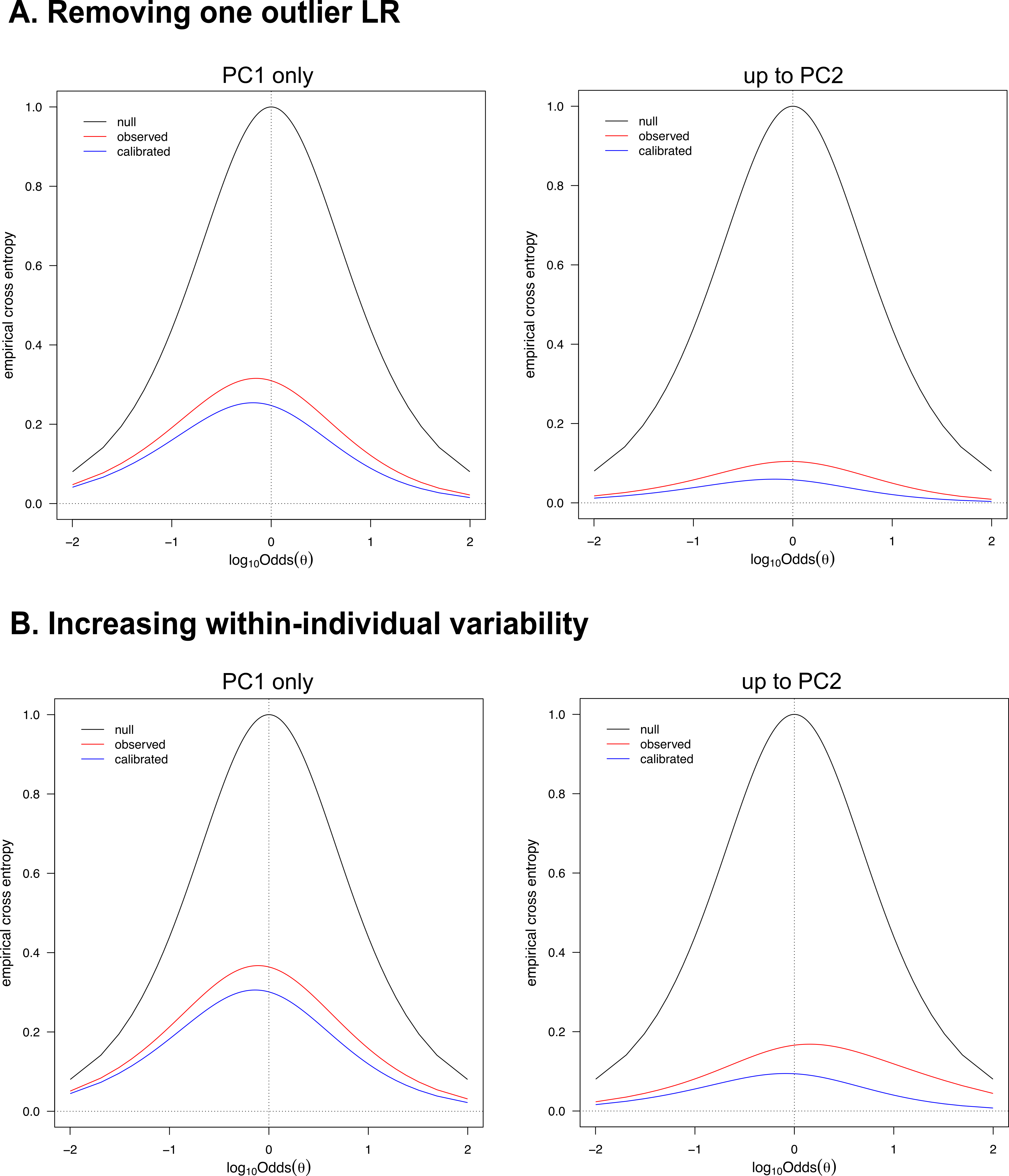}
    \caption{Empirical cross entropy plots of the likelihood ratios produced by the two-level model from scores on one or two PCs using footage from Dataset B, showing how removing one outlier LR or increasing within-individual variability (here to twice its original value) can improve the calibration.}
    \label{fig:ece_fix}
\end{figure}

Closer inspection reveals that this can be attributed to a single data point out of the total 54: on their third walk, Participant 14 showed two features of gait differently from the first two walks; using this walk as the query footage against the other two as reference produced an LR of $1e-3$. The leverage of this misleading LR increases as the prior odds favours the same-source proposition. Removing this LR greatly improves model calibration, \hl{illustrating that PCA can be very sensitive to outliers} (Figure \ref{fig:ece_fix}A). This misleadingly small LR indicates that the within-individual variability assumed by the model fits poorly to this particular case, where either the individual has higher variability compared to the general population, and/or the limited sampling of two pieces of reference footage failed to capture their entire range of variations. Indeed, when we increase the within-individual variability to twice its current value (the second row in Table \ref{tab:within_var}), the ECE curve generated using LRs from the entire dataset also remains below that of the null model (Figure \ref{fig:ece_fix}B).  

\subsubsection{Mis-specifying within-individual variability}\label{misspecify}

\begin{figure}[H]
    \centering
    \includegraphics[width=\textwidth]{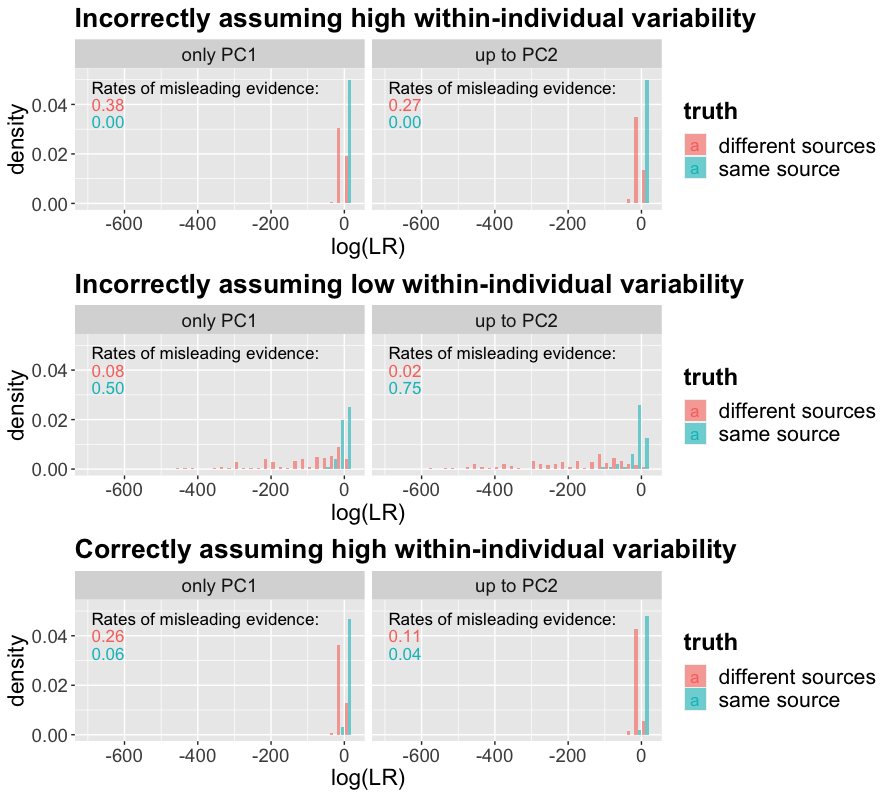}
    \caption{Histogram of log (base e) likelihood ratios obtained using a two-level model from scores on the first one or two PCs, showing the effect of mis-specifying the amount of within-individual variability. Low within-individual variability corresponds to estimates from Dataset B; high within-individual variability corresponds to estimates from Dataset A. The result from correctly assuming low within-individual variability is shown in Figure \ref{fig:lr_hist}. The rates of misleading evidence are the proportion of LRs being greater than one in different-source comparisons (pink), and the proportion of LRs being less than one in same-source comparisons (blue).}
    \label{fig:lr_mis}
\end{figure}

We explored the effect of mis-specifying the level of within-individual variability in the model, namely when we wrongly specify high variability within the same individual, as observed in our \textit{ad hoc} Dataset A, and when we wrongly specify limited variability within the same individual to \textit{ad hoc} observations (Figure \ref{fig:lr_mis}). Allowing high variability greatly reduces the magnitude of log LRs from different-source comparisons and causes some to fall on the misleading (greater than 1) side: the rate of misleading LRs is 0.27 when the first two PCs are used. The highest risk probably occurs when within-individual variation is actually high (in dataset A) but wrongly assumed to be low - for same-source comparisons, the log LR has the wrong sign in 50\% of cases starting with the first PC, and only worsens as the second PC is also considered. When the high within-individual variability is correctly modelled, however, dataset A also produces reasonable range of LRs. These results highlight the importance of properly assessing individual variability, as using the incorrect variance assumption leads to a large increase in misleading evidence. 

\hl{The variability within an individual does affect the conclusion as regards the weight of the evidence. This is why the more pieces of footage there are the more robust the conclusions are. A judgement can then be made as to how much and where variation is present within an individual and whether it arises from intrinsic or extrinsic factors. If the relative amount and type of intrinsic variability of the figure in the questioned footage is the same as the subject in the reference footage, the evidence will strengthen the hypothesis that the two individuals are the same. If the relative amount and type of variability differs between the two, the evidence will strengthen the hypothesis that the two individuals are different.}

\section{Discussion}

To explore the probative value of observational gait analysis, we examined the within- and between-individual variations in the features of gait assessed using the Sheffield Features of Gait Tool \cite{birch_repeatability_2019,birch_accuracy_2021}. The influence of demographic factors on features of gait suggests that what constitutes the relevant population needs to be carefully considered, regardless of whether a human expert or a statistical model is evaluating the evidence. We also observed high correlations between some features of gait, confirming that they should not contribute independently to the total weight of evidence.

High within-individual variability observed in the researchers' self-recordings motivated us to conduct a controlled study on recruited participants. When participants were asked to walk naturally along the same flat ground route in the same footwear with no handheld objects or heavy outerwear, and filmed from clear camera angles, there were minimal changes in individual features across three recording days. If this conclusion is reliable, the high variability we observed in the features of gait in Dataset A could be due to 1) uncontrolled factors in the self-recordings that alter the gait, such as garments, footwear, mood, walking speed, ground slope and material, or 2) factors that hinder reliable observation of the features of gait, such as lighting condition, frame rate and camera angle. All these factors are highly situational and changeable in actual casework; they are among the considerations when gait experts assess the suitability of footage for comparison, and when they summarise the result of comparison into an opinion. Using our LR model, we have seen that wrongly assuming high within-individual variability when it is low shifts different-source LRs closer to one; more dangerously, wrongly assuming low within-individual variability when it is high shifts most same-source LRs to smaller than one. This is not an uncommon scenario in practice considering the numerous factors at work. Therefore the current model at best provides assistance to gait experts' judgement; human expertise is indispensable in controlling the risk from a mis-specified model. Conducting empirical studies to assess the separate and joint effects of these factors on the observed features of gait is a worthwhile yet resource-intensive endeavour. On the other hand, experts of gait and video analysis usually have a solid understanding of these issues. A promising direction might be to elicit expert knowledge on factors that affect gait and sources of uncertainty, so that they can be incorporated into the statistical model. From the modelling perspective, this might require adjusting the estimate of within-individual variability according to the disparities in conditions between the questioned footage and the reference footage, or according to the uncertainty of the expert in assigning the features of gait. Statistical models could also help with the calibration of expert opinions, so that expert judgement, \hl{statistical} tools, and empirical data can work together to improve the evaluation of gait evidence.

The challenge of modelling the distribution of correlated ordinal or categorical variables is not unique to gait analysis, but applies to other forensic disciplines (e.g. comparison of facial images \cite{facial_identification_scientific_working_group_facial_2019} and footwear \cite{vernon_footwear_2017}) as well. The discrete categories could be the natural scale of outcome, or a result of human expert judgement. Developing a LR model for such variables would facilitate evaluative reporting of other types of evidence. We have recoded features of gait consisting of ordinal variables into binary variables, followed by dimension reduction to feed into a two-level KDE likelihood ratio model. Preliminary results are encouraging: under reasonable assumptions about within-individual variability, the rate of misleading evidence is below 0.1 for both same-source and different-source comparisons using up to four PCs. It came to our attention that including information from more PCs does not always increase the value of evidence, especially in same-source comparisons. We note that this could be due to using a population dataset with only one observation for each individual to establish the PC space; If variations within the same individual follow a different pattern from those between individuals, they could become compressed during dimension reduction and overlooked in downstream modelling. However, using the Dataset B which includes 3 observations for each individual did not eliminate the problem, although the sample size is likely too small for a firm conclusion. PCA seeks to retain the overall variations in the reduced dimensions. For the purpose of forensic comparison, ideally we want to find a sufficient dimension reduction with regard to preserving the LR, not maximising the overall variance. The problem could also exist in dimension reduction techniques that seek to find spaces that maximise between-group variation while minimising within-group variation, such as partial least squares discriminant analysis \cite{varmuza_introduction_2016}, regularised MANOVA and ANOVA simultaneous component analysis \cite{marini_analysis_2015}. These approaches followed by a score-based LR model have been applied in chemometric analysis in forensics \cite{allen_application_2019,martyna_forensic_2018}. It is not clear whether maximising between-group while minimising within-group variation preserves the most information relevant for calculating an LR - the consequence is likely to depend on the specific pattern of within- and between-group variations, the makeup of the dataset used in dimension reduction, as well as the subsequent modelling. A future research direction could be to investigate how various dimension reduction techniques affects the reliability of LRs to suggest the preferred practice for various types of evidence. 

We note that when tested on Dataset B, the calibration of our system is not fully satisfactory. This is caused by one or two misleadingly small LRs in same-source comparisons, where the individual performed one walk differently from the other two (However, not all such cases lead to LRs small enough to skew the ECE curve). Apart from a need to understand better how dimension reduction accentuates the differing features as mentioned above, it reflects that the assumption of equal within-individual variability along each PC, which we made considering the limited sample size in terms of both individuals and instances of walks, may not be appropriate. Collecting more empirical data will hopefully enable us to fit a model that accounts for non-constant variability between individuals, such as in \cite{bozza_probabilistic_2008}). We believe a better understanding of the distribution of individual variability and the sources of variation in features of gait is crucial towards developing a casework-ready model. We have shown in this work that it is possible and feasible to assign values to the numerator and denominator of an LR to evaluate probative value of gait observations in addressing source-level propositions about the person in the questioned footage.

\section*{Ethics statement}
The study to collect Dataset B has been approved by the School of Mathematics, University of Edinburgh ethics officer on 3 March 2022 (approval \#22.01).

\section*{Acknowledgement}
The authors thank Bert Otten and Mickey Wiedemeijer for sharing and discussing their model with us, and thank Gail Robertson and Marjan Sjerps for helpful discussions. The authors also thank all participants in our study for contributing their time. 

\section*{Funding}
This work was supported by Wave 1 of The UKRI Strategic Priorities Fund under the EPSRC Grant EP/W006022/1, particularly the “Criminal Justice System” theme within that grant and The Alan Turing Institute. 

\printbibliography

@Manual{r_psych,
    title = {psych: Procedures for Psychological, Psychometric, and Personality Research},
    author = {William Revelle},
    organization = { Northwestern University},
    address = { Evanston, Illinois},
    year = {2022},
    note = {R package version 2.2.5},
    url = {https://CRAN.R-project.org/package=psych},
  }

@Manual{R,
    title = {R: A Language and Environment for Statistical Computing},
    author = {{R Core Team}},
    organization = {R Foundation for Statistical Computing},
    address = {Vienna, Austria},
    year = {2022},
    url = {https://www.R-project.org/},
  }

@article{pearson1900,
 ISSN = {02643952},
 URL = {http://www.jstor.org/stable/90764},
 author = {Karl Pearson},
 journal = {Philosophical Transactions of the Royal Society of London. Series A, Containing Papers of a Mathematical or Physical Character},
 number = {},
 pages = {1--405},
 publisher = {The Royal Society},
 title = {Mathematical Contributions to the Theory of Evolution. VII. On the Correlation of Characters not Quantitatively Measurable},
 urldate = {2022-10-17},
 volume = {195},
 year = {1900}
}

@techreport{GroningenMethod,
    author = {Bert Otten and Mickey Wiedemeijer},
    title = {Forensic Gait Analysis Method Paper},
    institution = {University Medical Center Groningen, University of Groningen},
    contact = {egbert.otten@umcg.nl},
    note = {Contact: egbert.otten@umcg.nl},
    year = 2021
}

@misc{primer,
author={{The Royal Society and the Royal Society of Edinburgh}},
title={The use
of statistics
in legal
proceedings a primer for courts},
year={2020},
note={\url{https://royalsociety.org/-/media/about-us/programmes/science-and-law/science-and-law-statistics-primer.pdf}}
}

@misc{ICCA,
author={{The Inns of Court College of Advocacy and the Royal Statistical Society}},
year={2019},
title={Statistics and probability for advocates:
Understanding the use of statistical
evidence in courts and tribunals},
note=
{\url{https://rss.org.uk/RSS/media/File-library/Publications/ICCA-RSS-guide-version-6-branded-171019-REV03-designed-covers.pdf}}
}

@misc{FSR_LR,
author={{Forensic {S}cience {R}egulator}},
year={2021},
title={Forensic {S}cience {R}egulator
Codes of Practice and
Conduct: Development of Evaluative Opinions},
note={\url{https://assets.publishing.service.gov.uk/media/602407728fa8f5146f0769d9/FSR-C-118_Interpretation_Appendix_Issue_1__002_.pdf}}
}

@article{MORRISON2022100270,
title = {Advancing a paradigm shift in evaluation of forensic evidence: The rise of forensic data science},
journal = {Forensic Science International: Synergy},
volume = {5},
pages = {100270},
year = {2022},
issn = {2589-871X},
doi = {https://doi.org/10.1016/j.fsisyn.2022.100270},
url = {https://www.sciencedirect.com/science/article/pii/S2589871X22000559},
author = {Geoffrey Stewart Morrison}
}

@article{seckiner_forensic_2024,
	title = {Forensic interpretation framework for body and gait analysis: feature extraction, frequency and distinctiveness},
	volume = {56},
	issn = {0045-0618, 1834-562X},
	url = {https://www.tandfonline.com/doi/full/10.1080/00450618.2022.2161636},
	doi = {10.1080/00450618.2022.2161636},
	shorttitle = {Forensic interpretation framework for body and gait analysis},
	pages = {338--354},
	number = {4},
	journaltitle = {Australian Journal of Forensic Sciences},
	shortjournal = {Australian Journal of Forensic Sciences},
	author = {Seckiner, Dilan and Mallett, Xanthé and Roux, Claude and Gittelson, Simone and Maynard, Philip and Meuwly, Didier},
	urldate = {2024-12-28},
	date = {2024-07-03},
	langid = {english},
}

@article{champod_establishing_2004,
	title = {Establishing the most appropriate databases for addressing source level propositions},
	volume = {44},
	issn = {13550306},
	url = {https://linkinghub.elsevier.com/retrieve/pii/S1355030604717086},
	doi = {10.1016/S1355-0306(04)71708-6},
	pages = {153--164},
	number = {3},
	journaltitle = {Science \& Justice},
	shortjournal = {Science \& Justice},
	author = {Champod, C. and Evett, I.W. and Jackson, G.},
	urldate = {2023-12-11},
	date = {2004-07},
	langid = {english},
}

@article{ramos_informationtheoretical_2013,
	title = {Information‐Theoretical Assessment of the Performance of Likelihood Ratio Computation Methods},
	volume = {58},
	issn = {0022-1198, 1556-4029},
	url = {https://onlinelibrary.wiley.com/doi/10.1111/1556-4029.12233},
	doi = {10.1111/1556-4029.12233},
	pages = {1503--1518},
	number = {6},
	journaltitle = {Journal of Forensic Sciences},
	shortjournal = {Journal of Forensic Sciences},
	author = {Ramos, Daniel and Gonzalez‐Rodriguez, Joaquin and Zadora, Grzegorz and Aitken, Colin},
	urldate = {2023-10-23},
	date = {2013-11},
	langid = {english},
}

@book{varmuza_introduction_2016,
	edition = {1},
	title = {Introduction to Multivariate Statistical Analysis in Chemometrics},
	isbn = {978-0-429-14504-9},
	url = {https://www.taylorfrancis.com/books/9781420059496},
	publisher = {{CRC} Press},
	author = {Varmuza, Kurt and Filzmoser, Peter},
	urldate = {2022-10-07},
	date = {2016-04-19},
	langid = {english},
	doi = {10.1201/9781420059496},
}

@article{bozza_probabilistic_2008,
	title = {Probabilistic Evaluation of Handwriting Evidence: Likelihood Ratio for Authorship},
	volume = {57},
	issn = {0035-9254, 1467-9876},
	url = {https://academic.oup.com/jrsssc/article/57/3/329/7113415},
	doi = {10.1111/j.1467-9876.2007.00616.x},
	shorttitle = {Probabilistic Evaluation of Handwriting Evidence},
	pages = {329--341},
	number = {3},
	journaltitle = {Journal of the Royal Statistical Society Series C: Applied Statistics},
	author = {Bozza, Silvia and Taroni, Franco and Marquis, Raymond and Schmittbuhl, Matthieu},
	urldate = {2023-09-29},
	date = {2008-06-01},
	langid = {english},
}

@misc{kolenikov_use_2004,
	title = {The use of discrete data in {PCA}: theory, simulations, and applications to socioeconomic indices},
	url = {https://www.measureevaluation.org/resources/publications/wp-04-85/at_download/document},
	author = {Kolenikov, Stanislav and Angeles, Gustavo},
	date = {2004},
}

@book{jolliffe_principal_1986,
	location = {New York, {NY}},
	title = {Principal Component Analysis},
	isbn = {978-1-4757-1904-8},
	url = {http://link.springer.com/10.1007/978-1-4757-1904-8},
	series = {Springer Series in Statistics},
	publisher = {Springer New York},
	author = {Jolliffe, I. T.},
	urldate = {2022-11-09},
	date = {1986},
	doi = {10.1007/978-1-4757-1904-8},
}

@incollection{vernon_footwear_2017,
	edition = {2},
	title = {Footwear Examination and Analysis},
	isbn = {978-1-315-39502-9},
	booktitle = {Forensic Podiatry},
	publisher = {{CRC} Press},
	author = {Vernon, Denis Wesley and {DiMaggio}, John A.},
	date = {2017},
	note = {Num Pages: 34},
}

@misc{facial_identification_scientific_working_group_facial_2019,
	title = {Facial Comparison Overview and Methodology Guidelines v1.0},
	url = {https://fiswg.org/fiswg_facial_comparison_overview_and_methodology_guidelines_V1.0_20191025.pdf},
	author = {{Facial Identification Scientific Working Group}},
	urldate = {2022-10-18},
	date = {2019-10-25},
}

@article{marini_analysis_2015,
	title = {Analysis of variance of designed chromatographic data sets: The analysis of variance-target projection approach},
	volume = {1405},
	issn = {00219673},
	url = {https://linkinghub.elsevier.com/retrieve/pii/S0021967315007839},
	doi = {10.1016/j.chroma.2015.05.060},
	shorttitle = {Analysis of variance of designed chromatographic data sets},
	pages = {94--102},
	journaltitle = {Journal of Chromatography A},
	shortjournal = {Journal of Chromatography A},
	author = {Marini, Federico and de Beer, Dalene and Joubert, Elizabeth and Walczak, Beata},
	urldate = {2022-10-07},
	date = {2015-07},
	langid = {english},
}

@article{allen_application_2019,
	title = {Application of likelihood ratios and optimal decision thresholds in fire debris analysis based on a partial least squares discriminant analysis ({PLS}-{DA}) model},
	volume = {16},
	issn = {24681709},
	url = {https://linkinghub.elsevier.com/retrieve/pii/S2468170919300852},
	doi = {10.1016/j.forc.2019.100188},
	pages = {100188},
	journaltitle = {Forensic Chemistry},
	shortjournal = {Forensic Chemistry},
	author = {Allen, Alyssa and Williams, Mary R. and Sigman, Michael E.},
	urldate = {2022-10-07},
	date = {2019-12},
	langid = {english},
}

@article{martyna_forensic_2018,
	title = {Forensic comparison of pyrograms using score-based likelihood ratios},
	volume = {133},
	issn = {01652370},
	url = {https://linkinghub.elsevier.com/retrieve/pii/S016523701830041X},
	doi = {10.1016/j.jaap.2018.03.024},
	pages = {198--215},
	journaltitle = {Journal of Analytical and Applied Pyrolysis},
	shortjournal = {Journal of Analytical and Applied Pyrolysis},
	author = {Martyna, Agnieszka and Zadora, Grzegorz and Ramos, Daniel},
	urldate = {2022-10-07},
	date = {2018-08},
	langid = {english},
}

@article{aitken_evaluation_2004,
	title = {Evaluation of Trace Evidence in the Form of Multivariate Data},
	volume = {53},
	issn = {0035-9254},
	url = {https://www.jstor.org/stable/3592690},
	pages = {109--122},
	number = {1},
	journaltitle = {Journal of the Royal Statistical Society. Series C (Applied Statistics)},
	author = {Aitken, C. G. G. and Lucy, D.},
	urldate = {2022-10-06},
	date = {2004},
	note = {Publisher: [Wiley, Royal Statistical Society]},
}

@article{connor_biometric_2018,
	title = {Biometric recognition by gait: A survey of modalities and features},
	volume = {167},
	issn = {1077-3142},
	url = {https://www.sciencedirect.com/science/article/pii/S1077314218300079},
	doi = {10.1016/j.cviu.2018.01.007},
	shorttitle = {Biometric recognition by gait},
	pages = {1--27},
	journaltitle = {Computer Vision and Image Understanding},
	shortjournal = {Computer Vision and Image Understanding},
	author = {Connor, Patrick and Ross, Arun},
	urldate = {2022-10-03},
	date = {2018-02-01},
	langid = {english},
}

@misc{european_network_of_forensic_science_institutes_enfsi_2016,
	title = {{ENFSI} Guideline for Evaluative Reporting in Forensic Science},
	url = {https://enfsi.eu/about-enfsi/structure/working-groups/documents-page/documents/forensic-guidelines/},
	author = {{European Network of Forensic Science Institutes}},
	urldate = {2022-09-29},
	date = {2016-09-27},
}

@article{evett_logical_2015,
	title = {The logical foundations of forensic science: towards reliable knowledge},
	volume = {370},
	issn = {0962-8436, 1471-2970},
	url = {https://royalsocietypublishing.org/doi/10.1098/rstb.2014.0263},
	doi = {10.1098/rstb.2014.0263},
	shorttitle = {The logical foundations of forensic science},
	pages = {20140263},
	number = {1674},
	journaltitle = {Philosophical Transactions of the Royal Society B: Biological Sciences},
	shortjournal = {Phil. Trans. R. Soc. B},
	author = {Evett, Ian},
	urldate = {2022-09-29},
	date = {2015-08-05},
	langid = {english},
}

@article{bergsma_foot_2021,
	title = {Foot placement variables of pedestrians in community setting during curve walking},
	volume = {86},
	issn = {09666362},
	url = {https://linkinghub.elsevier.com/retrieve/pii/S0966636221001028},
	doi = {10.1016/j.gaitpost.2021.03.017},
	pages = {120--124},
	journaltitle = {Gait \& Posture},
	shortjournal = {Gait \& Posture},
	author = {Bergsma, B. and Hulleman, D.N. and Wiedemeijer, M.M. and Otten, E.},
	urldate = {2022-09-29},
	date = {2021-05},
	langid = {english},
}

@incollection{jackson_probative_2020,
	title = {Probative value of gait analysis},
	isbn = {978-0-429-42658-2},
	booktitle = {Forensic Gait Analysis},
	publisher = {{CRC} Press},
	author = {Jackson, Graham and Birch, Ivan},
	date = {2020},
	note = {Num Pages: 18},
}

@article{macoveciuc_forensic_2019,
	title = {Forensic Gait Analysis and Recognition: Standards of Evidence Admissibility},
	volume = {64},
	issn = {1556-4029},
	url = {https://onlinelibrary.wiley.com/doi/abs/10.1111/1556-4029.14036},
	doi = {10.1111/1556-4029.14036},
	shorttitle = {Forensic Gait Analysis and Recognition},
	pages = {1294--1303},
	number = {5},
	journaltitle = {Journal of Forensic Sciences},
	author = {Macoveciuc, Ioana and Rando, Carolyn J. and Borrion, Hervé},
	urldate = {2022-09-28},
	date = {2019},
	langid = {english},
	note = {\_eprint: https://onlinelibrary.wiley.com/doi/pdf/10.1111/1556-4029.14036},
}

@thesis{seckiner_development_2021,
	title = {The Development and Testing of a Forensic Interpretation Framework for use on Anthropometric and Morphological Data Collected During Stance and Gait},
	url = {https://opus.lib.uts.edu.au/handle/10453/151212},
	institution = {University of Technology Sydney},
	type = {phdthesis},
	author = {Seckiner, Dilan},
	date = {2021},
}

@article{wiedemeijer_effects_2018,
	title = {Effects of high heeled shoes on gait. A review},
	volume = {61},
	issn = {0966-6362},
	url = {https://www.sciencedirect.com/science/article/pii/S0966636218300687},
	doi = {10.1016/j.gaitpost.2018.01.036},
	pages = {423--430},
	journaltitle = {Gait \& Posture},
	shortjournal = {Gait \& Posture},
	author = {Wiedemeijer, M. M. and Otten, E.},
	urldate = {2022-09-28},
	date = {2018-03-01},
	langid = {english},
}

@article{reidy_effect_2020,
	title = {The effect of viewing angle on observations of foot orientation in forensic gait analysis},
	volume = {60},
	issn = {1355-0306},
	url = {https://www.sciencedirect.com/science/article/pii/S1355030620300642},
	doi = {10.1016/j.scijus.2020.06.005},
	pages = {504--511},
	number = {6},
	journaltitle = {Science \& Justice},
	shortjournal = {Science \& Justice},
	author = {Reidy, Selina and Stephenson, John and Smith, Francine and Otten, Egbert and Wiedemeijer, Mickey and Curran, Mike},
	urldate = {2022-09-28},
	date = {2020-11-01},
	langid = {english},
}

@book{abboud_forensic_2017,
	title = {Forensic Gait Analysis: a primer for courts},
	isbn = {978-1-78252-302-4},
	url = {https://royalsociety.org/about-us/programmes/science-and-law/},
	series = {Primers for courts},
	author = {Abboud, Rami and Baker, Richard and Stebbins, Julie and Wall, Mark and Black, Sue},
	editor = {Cubie, Andrew and Theologis, Tim and Wolpert, Daniel},
	urldate = {2022-09-28},
	date = {2017},
}

@article{van_mastrigt_critical_2018,
	title = {Critical review of the use and scientific basis of forensic gait analysis},
	volume = {3},
	issn = {2096-1790, 2471-1411},
	url = {https://www.tandfonline.com/doi/full/10.1080/20961790.2018.1503579},
	doi = {10.1080/20961790.2018.1503579},
	pages = {183--193},
	number = {3},
	journaltitle = {Forensic Sciences Research},
	shortjournal = {Forensic Sciences Research},
	author = {van Mastrigt, Nina M. and Celie, Kevin and Mieremet, Arjan L. and Ruifrok, Arnout C. C. and Geradts, Zeno},
	urldate = {2022-09-26},
	date = {2018-07-03},
	langid = {english},
}

@article{nirenberg_review_2018,
	title = {A review of the historical use and criticisms of gait analysis evidence},
	volume = {58},
	issn = {1355-0306},
	url = {https://www.sciencedirect.com/science/article/pii/S1355030617301788},
	doi = {10.1016/j.scijus.2018.03.002},
	pages = {292--298},
	number = {4},
	journaltitle = {Science \& Justice},
	shortjournal = {Science \& Justice},
	author = {Nirenberg, Michael and Vernon, Wesley and Birch, Ivan},
	urldate = {2022-04-26},
	date = {2018-07-01},
	langid = {english},
}

@article{silverman_choosing_1978,
	title = {Choosing the window width when estimating a density},
	volume = {65},
	issn = {0006-3444, 1464-3510},
	url = {https://academic.oup.com/biomet/article-lookup/doi/10.1093/biomet/65.1.1},
	doi = {10.1093/biomet/65.1.1},
	pages = {1--11},
	number = {1},
	journaltitle = {Biometrika},
	shortjournal = {Biometrika},
	author = {Silverman, B. W.},
	urldate = {2022-08-17},
	date = {1978},
	langid = {english},
}

@article{birch_aiding_2016,
	title = {Aiding the interpretation of forensic gait analysis: Development of a features of gait database},
	volume = {56},
	issn = {1355-0306},
	url = {https://www.sciencedirect.com/science/article/pii/S1355030616300570},
	doi = {10.1016/j.scijus.2016.06.009},
	shorttitle = {Aiding the interpretation of forensic gait analysis},
	pages = {426--430},
	number = {6},
	journaltitle = {Science \& Justice},
	shortjournal = {Science \& Justice},
	author = {Birch, Ivan and Gwinnett, Claire and Walker, Jeremy},
	urldate = {2022-04-27},
	date = {2016-12-01},
	langid = {english},
}

@article{birch_development_2013,
	title = {The development of a tool for assessing the quality of closed circuit camera footage for use in forensic gait analysis},
	volume = {20},
	issn = {1752-928X},
	url = {https://www.sciencedirect.com/science/article/pii/S1752928X13002126},
	doi = {10.1016/j.jflm.2013.07.005},
	pages = {915--917},
	number = {7},
	journaltitle = {Journal of Forensic and Legal Medicine},
	shortjournal = {Journal of Forensic and Legal Medicine},
	author = {Birch, Ivan and Vernon, Wesley and Walker, Jeremy and Saxelby, Jai},
	urldate = {2022-04-26},
	date = {2013-10-01},
	langid = {english},
}

@article{birch_accuracy_2021,
	title = {The accuracy and validity of the Sheffield Features of Gait Tool},
	volume = {61},
	issn = {1355-0306},
	url = {https://www.sciencedirect.com/science/article/pii/S1355030620302574},
	doi = {10.1016/j.scijus.2020.08.001},
	pages = {72--78},
	number = {1},
	journaltitle = {Science \& Justice},
	shortjournal = {Science \& Justice},
	author = {Birch, Ivan and Birch, Maria and Lall, Jalmeen},
	urldate = {2022-04-26},
	date = {2021-01-01},
	langid = {english},
}

@article{birch_repeatability_2019,
	title = {The repeatability and reproducibility of the Sheffield Features of Gait Tool},
	volume = {59},
	issn = {1355-0306},
	url = {https://www.sciencedirect.com/science/article/pii/S1355030618303563},
	doi = {10.1016/j.scijus.2019.04.001},
	pages = {544--551},
	number = {5},
	journaltitle = {Science \& Justice},
	shortjournal = {Science \& Justice},
	author = {Birch, Ivan and Birch, Maria and Rutler, Lucy and Brown, Sarah and Burgos, Libertad Rodriguez and Otten, Bert and Wiedemeijer, Mickey},
	urldate = {2022-04-26},
	date = {2019-09-01},
	langid = {english},
}

\end{document}